\documentclass[sn-mathphys-num]{sn-jnl}

\usepackage{graphicx}%
\usepackage{multirow}%
\usepackage{amsmath,amssymb,amsfonts}%
\usepackage{amsthm}%
\usepackage{mathrsfs}%
\usepackage[title]{appendix}%
\usepackage{xcolor}%
\usepackage{textcomp}%
\usepackage{manyfoot}%
\usepackage{booktabs}%
\usepackage{algorithm}%
\usepackage{algorithmicx}%
\usepackage{algpseudocode}%
\usepackage{listings}%

\theoremstyle{thmstyleone}%
\theoremstyle{thmstyletwo}%

\theoremstyle{thmstylethree}%

\begin{document}

\title[Fluid conductivity in porous media in terms of beta functions]{Fluid conductivity in porous media in terms of beta functions}


\author*[1]{\fnm{Tairone} \spfx{Paiva} \sur{Le\~{a}o}}\email{tleao@unb.br}

\affil*[1]{\orgdiv{Soil Physics Laboratory, Faculty of Agronomy and Veterinary Medicine}, \orgname{University of Brasilia}, \orgaddress{\street{Campus Universit\'{a}rio Darcy Ribeiro}, \city{Brasilia}, \postcode{70910-900}, \state{Distrito Federal}, \country{Brazil}}}


\abstract{Conductivity of unsaturated porous media to fluids is of theoretical and applied interest to mathematicians, physicists, and chemical, petroleum, civil and agricultural engineers. We explore the expression of unsaturated relative conductivity equations as a regularized incomplete beta function, which implies a beta distribution of the variable of interest. A generalized solution for $n>1$ and $m>0$ is presented. The solution can be implemented without difficulties in most programming languages and simulation software.}




\maketitle


\section{Introduction}\label{sec1}

Childs and Collis-George (1950), Burdine (1953) and Millington and Quirk (1961) proposed a theoretical framework in which the relative permeability of unsaturated soil can be described in terms of unknown pore distribution functions. In its most essential form, the theory states that pores of radius $\sigma$ with distribution $f(\sigma)$ at a plane $y-z$ located at $x$ encounter pores of radius $\rho$ with distribution $f(\rho)$ at $x + \Delta x$. If $f(\sigma)d\sigma$ indicate the fraction of the area in $y-z$ occupied by pores of radius $\sigma + \delta \sigma$ and $f(\rho)d\rho$ indicate the fraction occupied by pores of radius $\rho + \rho d\rho$, then the probability of pores of radius $\sigma + \delta \sigma$ encountering pores of radius $\rho + \rho d\rho$ at $x + \delta x$ is (Childs and Collis-George 1950; Mualem 1976) 
\begin{equation}
a(\sigma, \rho) = f(\sigma)d\sigma f(\rho)d\rho
\label{eq1}
\end{equation}
The model was generalized for approximating real porous medium as (Mualem 1976)
\begin{equation}
a(\sigma, \rho) = G(R, \sigma, \rho) T(R, \sigma, \rho) f(\sigma)d\sigma f(\rho)d\rho
\label{eq2}
\end{equation}
In which $G$ and $T$ are unknown correction factors accounting for partial correlation between pores of distributions  $f(\sigma)$ and $f(\rho)$ and tortuosity. The relative permeability $K_r$ was to be found by the ratio of integrals of Eq. (\ref{eq2}) from zero to the pores of maximum radius filled with water $R$ in the numerator and from zero to the maximum pore radius in the denominator $R_{max}$. Mualem (1976) also replaced the pore radius distribution with a water potential-water content function using Laplace's equation resulting in  
\begin{equation}
K_r = \Theta^{\gamma} [ \int_0^\Theta \frac{1}{\psi(\Theta)} d \Theta / \int_0^1 \frac{1}{\psi(\Theta)} d \Theta]^2
\label{eq3}
\end{equation}
The role of the $\Theta^{\gamma}$ to act as a correction term accounting for the unknown effects of tortuosity and partial correlation. The value of $\gamma = 1/2$ was adopted, being an empirical value.     

Several water retention functions were used as distribution functions such as the Brooks and Corey (1964), Haverkamp et al. (1977) and later the van Gentuchten (1980) equation. The van Genuchten equation provides an interesting case study because of its relationship to special functions and is investigated in this study.   

\section{Theoretical development}\label{sec2}

Let $\theta(\psi)$ be a semi-empirical function which describes the water retention curve for a porous medium (van Genuchten 1980) 

\begin{equation}
\theta = \theta_r + (\theta_s - \theta_r) (1 + (\alpha \psi)^{n})^{-m}
\label{eq4}
\end{equation}
with
\begin{equation}
\Theta = \frac{(\theta - \theta_r)}{(\theta_s - \theta_r)}
\label{eq5}
\end{equation}
the normalized form of $\theta$ with $\Theta \in [0,1]$, also known as effective saturation. Following the procedure of van Genuchten (1980), a dimensionless conductivity function can be obtained from Eq. (\ref{eq3}) by solving Eq. (\ref{eq4}) for $\psi$   

\begin{equation}
\psi(\Theta) =  \frac{(\Theta^{-\frac{1}{m}}-1)^{\frac{1}{n}}}{\alpha} 
\label{eq6}
\end{equation}
and defining the numerator and denominator integrals in Eq. (\ref{eq3}) as 

\begin{equation}
C =  \int_0^\Theta \frac{1}{\psi(\Theta)} d \Theta 
\label{eq7}
\end{equation}
\begin{equation}
D =  \int_0^1 \frac{1}{\psi(\Theta)} d \Theta
\label{eq8}
\end{equation}
respectively. By using the substitution $\Theta = \zeta^m$, $d \Theta = m \zeta^{m-1} d \zeta$, Eq. (\ref{eq6}) is written as
\begin{equation}
\psi(\zeta) =  \frac{(\zeta^{-1}-1)^{\frac{1}{n}}}{\alpha} 
\label{eq9}
\end{equation}
such that
\begin{equation}
C =  \alpha m \int_0^{\zeta} \zeta^{\frac{1}{n}+m-1}( 1 - \zeta)^{-\frac{1}{n}} d \zeta
\label{eq10}
\end{equation}
\begin{equation}
D =  \alpha m \int_0^1 \zeta^{\frac{1}{n}+m-1}( 1 - \zeta)^{-\frac{1}{n}} d \zeta
\label{eq11}
\end{equation}
Without loss of generality, we can write 
\begin{equation}
-\frac{1}{n} = 1 - \frac{1}{n} - 1
\label{eq12}
\end{equation}
and considering that $\alpha m$ will appear in the numerator and denominator in the inner term of Eq. (\ref{eq3}), $C$ and $D$ can be written as
\begin{equation}
B(\zeta; \frac{1}{n}+m, 1-\frac{1}{n}) =  \int_0^{\zeta} \zeta^{\frac{1}{n}+m-1}( 1 - \zeta)^{1-\frac{1}{n}-1} d \zeta
\label{eq13}
\end{equation}
\begin{equation}
B(\frac{1}{n}+m, 1-\frac{1}{n}) =  \int_0^1 \zeta^{\frac{1}{n}+m-1}( 1 - \zeta)^{1-\frac{1}{n}-1} d \zeta
\label{eq14}
\end{equation}
the incomplete and complete beta functions, respectively (Abramovitz and Stegun, 1965). Now, the relative hydraulic conductivity is written as 
\begin{equation}
K_r = \zeta^{\frac{m}{2}} [\frac{B(\zeta; \frac{1}{n}+m, 1-\frac{1}{n})}{B(\frac{1}{n}+m, 1-\frac{1}{n})} ]^2
\label{eq15}
\end{equation}
Restoring the original variable
\begin{equation}
K_r = \Theta^{\frac{1}{2}} [\frac{B(\Theta^{\frac{1}{m}}; \frac{1}{n}+m, 1-\frac{1}{n})}{B(\frac{1}{n}+m, 1-\frac{1}{n})} ]^2
\label{eq16}
\end{equation}
But 
\begin{equation}
I(\Theta^{\frac{1}{m}}; \frac{1}{n}+m, 1-\frac{1}{n})=\frac{B(\Theta^{\frac{1}{m}}; \frac{1}{n}+m, 1-\frac{1}{n})}{B(\frac{1}{n}+m, 1-\frac{1}{n})} 
\label{eq17}
\end{equation}
defines the regularized incomplete beta function (Abramovitz and Stegun, 1965), such that
\begin{equation}
K_r = \Theta^{\frac{1}{2}} [I(\Theta^{\frac{1}{m}}; \frac{1}{n}+m, 1-\frac{1}{n}) ]^2
\label{eq18}
\end{equation}
which, in its more general form, is valid for $m > 0$ and $n > 1$. A specific case is when
\begin{equation}
m = 1-\frac{1}{n}
\label{eq19}
\end{equation}
restricting $m \in (0,1)$ and resulting 
\begin{equation}
K_r = \Theta^{\frac{1}{2}} [I(\Theta^{\frac{1}{m}}; 1, m) ]^2
\label{eq20}
\end{equation}
By the properties of the incomplete beta function
\begin{equation}
I(x; 1, b) = 1 - (1-x)^b
\label{eq21}
\end{equation}
\emph{Proof:}
\begin{equation}
I(x; a, b) = \frac{B_x(x; a, b)}{B(a, b)} = \frac{\int_0^x t^{a-1} (1-t)^{b-1} dt}{\int_0^1 t^{a-1} (1-t)^{b-1} dt}  
\label{eq22}
\end{equation}
\begin{equation}
I(x; 1, m) = \frac{B_x(x; 1, m)}{B(1, m)} = \frac{\int_0^x (1-t)^{b-1} dt}{\int_0^1  (1-t)^{b-1} dt} = 1 - (1-x)^{m} 
\label{eq23}
\end{equation}
Thus, in analytical form, the special case of Eq. (\ref{eq20}) is
\begin{equation}
K_r = \Theta^{\frac{1}{2}} [1 - (1 - \Theta^{\frac{1}{m}})^m) ]^2
\label{eq24}
\end{equation}
This is the solution found by van Genuchten (1980) by direct integration of Eqs. (\ref{eq10}) and (\ref{eq11}) under the condition given by Eq. (\ref{eq19}). 

The regularized incomplete beta function is the cumulative probability of the beta distribution (Guthrie, 2020), thus
\begin{equation}
F(x; a, b) = I(x; a, b)
\label{eq25}
\end{equation}
with moments 
\begin{equation}
E[X] = \frac{a}{a+b}
\label{eq26}
\end{equation}
\begin{equation}
var[X] = \frac{ab}{(a+b)^2 (a + b + 1) }
\label{eq27}
\end{equation}
Thus, the inner term of the conductivity function is a cumulative statistical distribution with mean and standard deviation 
\begin{equation}
\mu = \frac{\frac{1}{n}+m}{m+1}
\label{eq28}
\end{equation}
\begin{equation}
\sigma^2 = \frac{(\frac{1}{n}+m)(1-\frac{1}{n})}{(m+1)^2 (m+2) }
\label{eq29}
\end{equation}
From Eq. (\ref{eq18})
\begin{equation}
\frac{K_r^{\frac{1}{2}}}{\Theta^{\frac{1}{4}}} = F(\Theta^{\frac{1}{m}}; \frac{1}{n}+m, 1-\frac{1}{n}) 
\label{eq30}
\end{equation}
with average $\mu$ and variance $\sigma^2$. Thus, the van Genuchten equation implies a beta distribution for the relative permeability. This is in accordance with early theory which was based on statistical assumptions regarding pore size distributions. 

In addition, the expression of permeability in terms of beta functions allows determination of permeability for different solutions of the van Genuchten integrals without the necessity of having to provide analytical solutions which can be complicated or nonexistent. For example, van Genuchten solution for $m=2-1/n$ is readily expressed as  
\begin{equation}
K_r = \Theta^{\frac{1}{2}} [I(\Theta^{\frac{1}{m}}; 2, m-1) ]^2
\label{eq31}
\end{equation}
which can be generalized for all positive integer $k=m-1+1/n$ as 
\begin{equation}
K_r = \Theta^{\frac{1}{2}} [I(\Theta^{\frac{1}{m}}; 1+k, m-k) ]^2
\label{eq32}
\end{equation} 
Another particular case is when both coefficient of the beta function are equal and in this case $1/n + m = 1 - 1/n$ and
\begin{equation}
K_r = \Theta^{\frac{1}{2}} [I(\Theta^{\frac{1}{m}}; 1/n + m, 1/n + m) ]^2
\label{eq33}
\end{equation} 
although using $1- 1/n$ is equally valid.

Equations expressed in terms of special functions can be fit to data without specific assumptions regarding the dependence between  $m$ and $n$. Although not being a closed-form, they can be fit to data using statistical techniques since special functions are now ubiquitous in statistical software and programming languages. Because of that they can also be used directly in simulation codes and numerical calculations for a range of conditions without the necessity of an analytical solution.


\section{Analysis and application}\label{sec2}

\begin{figure}
\centering
 \includegraphics[width=0.75\textwidth]{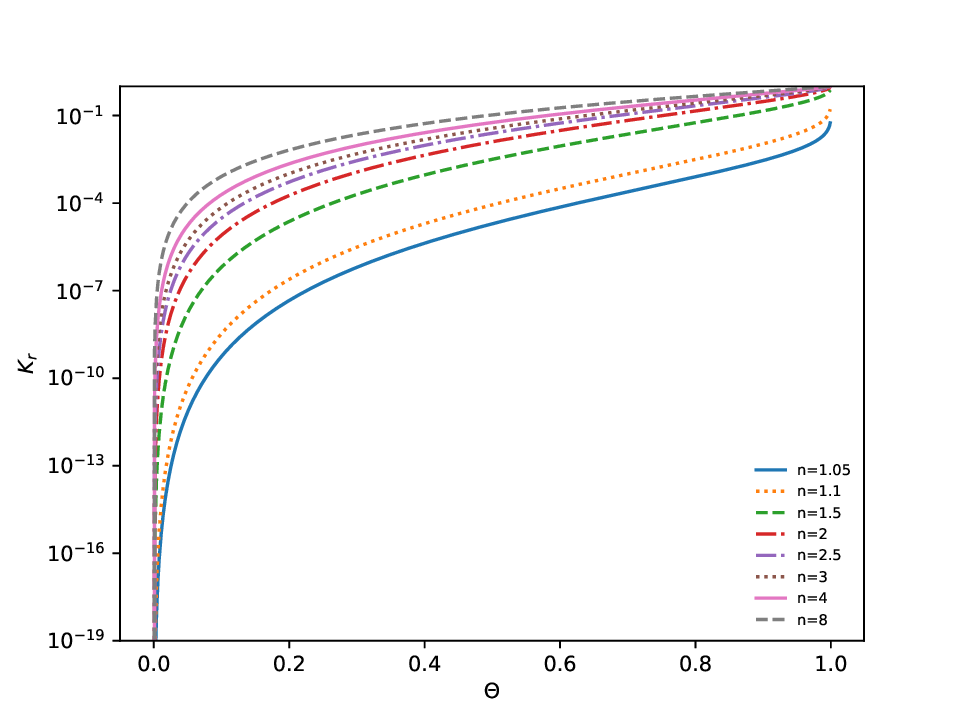}
\caption{Relative conductivity as a function of relative saturation for $m=0.5$. }
\label{fig1}
\end{figure}

\begin{figure}
\centering
 \includegraphics[width=0.75\textwidth]{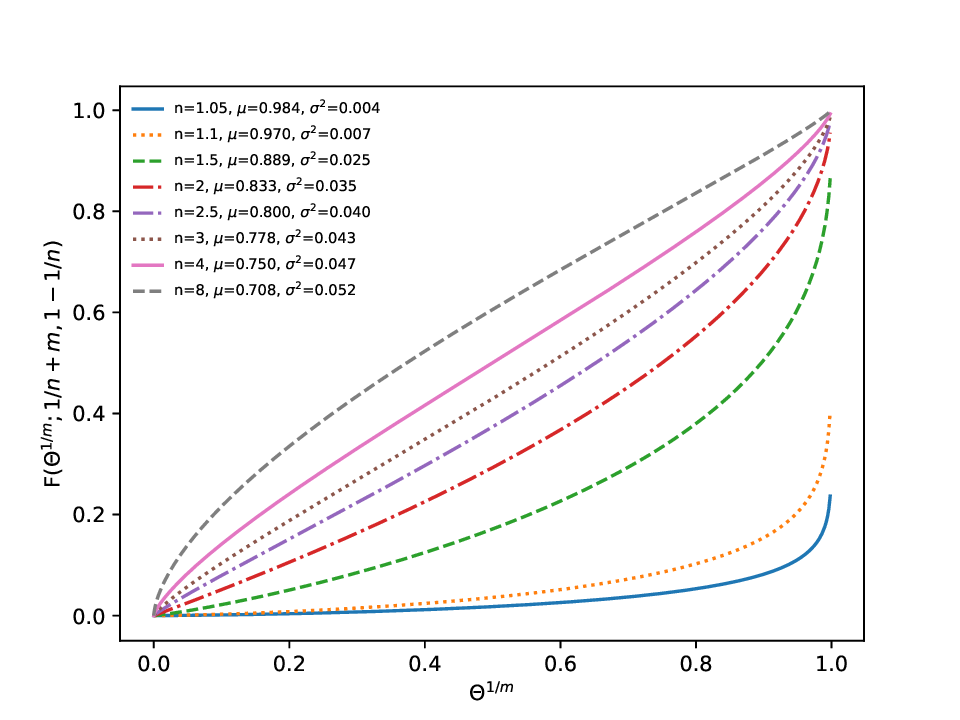}
\caption{Cumulative beta distribution for $m=0.5$. }
\label{fig2}
\end{figure}

The distribution of $K_r$ values calculated from Eq. (\ref{eq18}) for $m=0.5$ and $n$ ranging from 1.05 to 8 is presented in Fig. \ref{fig1}. The line for the analytical solution with $m=1-1/n$ is indicated. The corresponding cumulative beta distribution is presented in Fig. \ref{fig2}. For $m=0.5$, the increase in $n$ decreases the average, $\mu$, while the variance $\sigma^2$ is maximum for $n=4$. As the average increases, the distribution will concentrate around higher $\Theta^{1/m}$ values.

\begin{figure}[H]
\centering
 \includegraphics[width=0.75\textwidth]{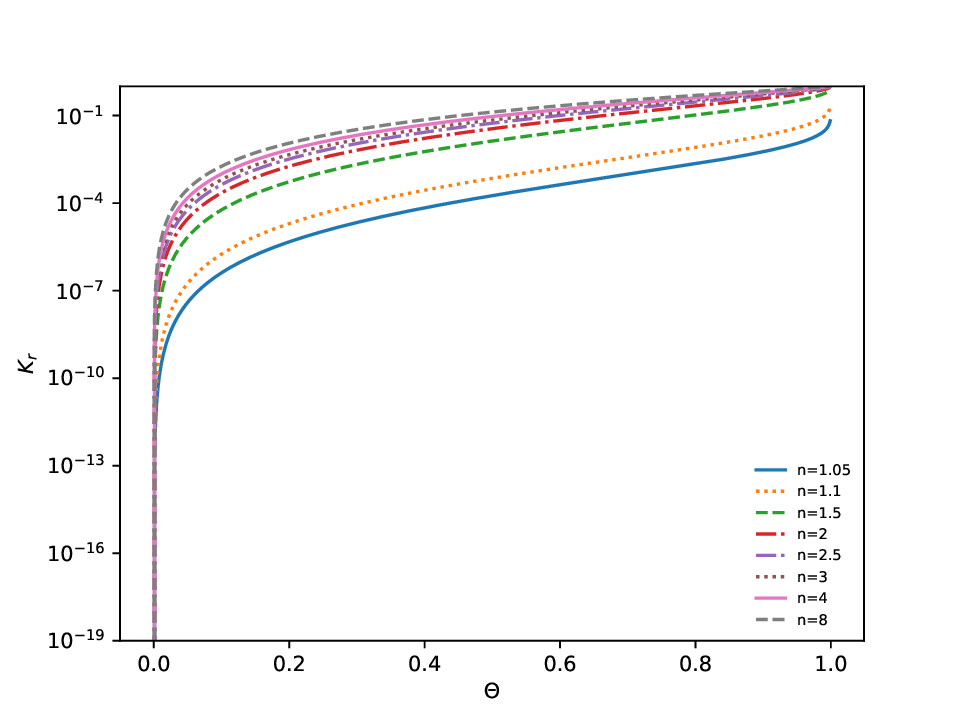}
\caption{Relative conductivity as a function of relative saturation for $m=4.0$. }
\label{fig3}
\end{figure}

\begin{figure}
\centering
 \includegraphics[width=0.75\textwidth]{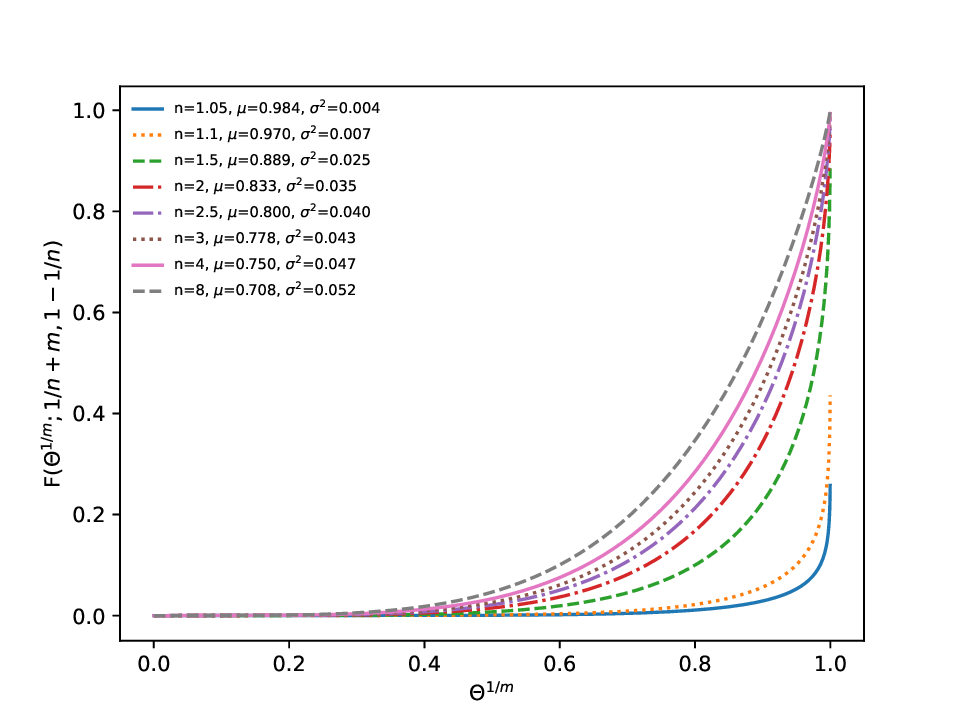}
\caption{Cumulative beta distribution for $m=4.0$. }
\label{fig4}
\end{figure}

Increasing to $m=4$ overall shifts the predictive curves to higher values of $K_r$ (Fig. \ref{fig3}). This causes an overall increase in the averages of the cumulative beta distributions (Fig. \ref{fig4}). As $n \to \infty$, Eq. (\ref{eq17}) becomes
\begin{equation}
I(x; m, 1) = x^{m} 
\label{eq34}
\end{equation}
which can be easily proved by arguments similar to used in Eq. (\ref{eq22}) to Eq. (\ref{eq23}). Under these conditions, the conductivity function, Eq. (\ref{eq18}), tends to  
\begin{equation}
K_r = \Theta^{\frac{5}{2}}
\label{eq35}
\end{equation}
which, incidentally, is the same result one would obtain from the conductivity function based on Brooks and Corey (1964) derived by van Genuchten (1980) from Eq. (\ref{eq3}), by taking the $\lambda$ pore size distribution index to infinity. As discussed by Brooks and Corey (1964) $\lambda$ is small for materials with a wide range of pores and large for uniform pore distribution, thus, for a material with uniform pore size distribution, the relative conductivity is a function of the relative saturation only. This shows that the $n$ parameter from van Genuchten (1980) has an analogous role as $\lambda$ from Brooks and Corey (1964). The same relationship between $K_r$ and $\Theta$ can be found by replacing $m=1-1/n$ in Eq. (\ref{eq24}) and making $n \to \infty$.


\begin{figure}
\centering
 \includegraphics[width=0.75\textwidth]{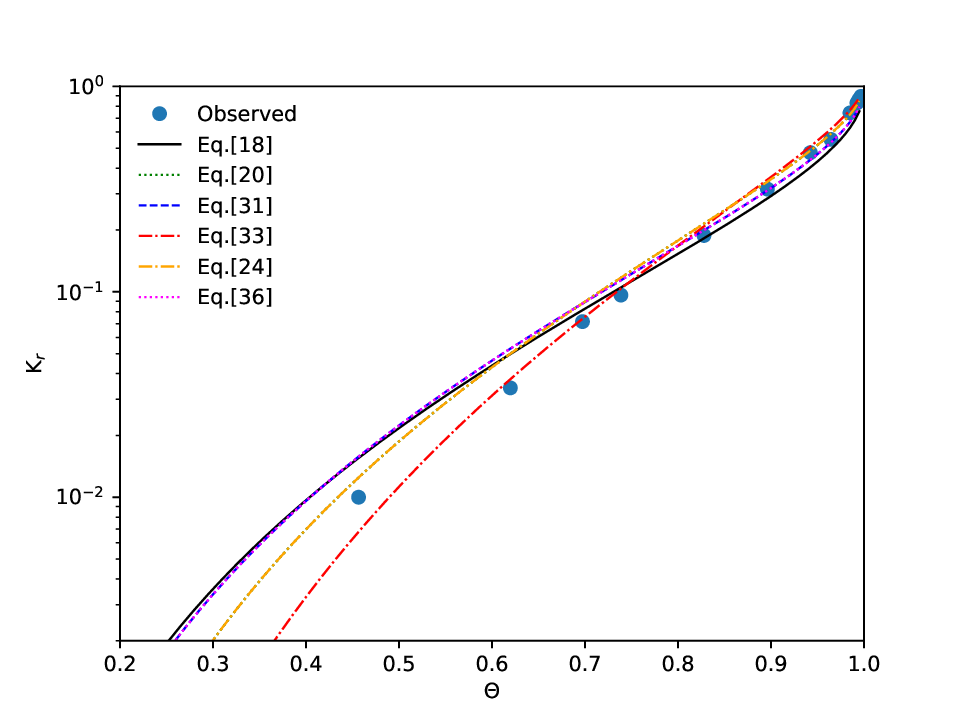}
\caption{Observed and estimated relative permeability as a function of relative saturation for selected models with parameters estimated for the water retention curve.}
\label{fig5}
\end{figure}

Water retention and permeability data from the ``Silt Loam G.E.3'' soil from van Genuchten (1980) was used to test predictive capability of the equations based on special functions and analytical solutions. Water potential was converted into relative saturation using the water retention curve. Values used in the predictions were $\theta_s=0.3954$, $\theta_r=0.1812$, $\alpha=0.00145$, $m=2.9705$ and $n=1.7145$, obtained from nonlinear least-squares of the $\theta(\psi)$ function. The only reason for using this fit for calculating relative saturation is because it provided the best fit for the water retention curve in comparison to other possibilities. Then values of $m, n$ calculated from the water retention function for each specific restriction, i.e. $m, n$ free parameters (Eq. (\ref{eq18})), $m=1-1/n$ (Eq. (\ref{eq20})), $m=2-1/n$ (Eq. (\ref{eq31})), $1/n+m = 1-1/n$ (Eq. (\ref{eq33})) were inserted into the special function equations and analytical solutions for $m=1-1/n$ (Eq. (\ref{eq24})) along with van Genuchten's solution for $m=2-1/n$ 
\begin{equation}
K_r = \Theta^{\frac{1}{2}} [1 - m(1 - \Theta^{\frac{1}{m}})^{(m-1)} + (m-1)(1 - \Theta^{\frac{1}{m}})^m) ]^2
\label{eq36}
\end{equation}
and used to predict relative permeability (Figure \ref{fig5}). As expected, the variability in the values of the parameters $m$ and $n$ influenced the shape and curvature of the predictions. Because the permeability data is obtained independently from the water retention curve, the predictive capability is not directly related to the goodness of fit of the water retention curve. The special function for $m=1-1/n$ (Eq. (\ref{eq20})) provided the best agreement with measured relative permittivity values, with Root Mean Square Error $RMSE = 0.0218$, followed by Eq.[\ref{eq33}] $RMSE = 0.0279$. For the other models $RMSE$ was $0.0464$, and $0.0738$ for Eq. (\ref{eq31}) and  Eq. (\ref{eq18}), respectively.  Not surprisingly, the predictions from the numerical and analytical solutions for $m=1-1/n$ (Eq. (\ref{eq20}) and Eq. (\ref{eq24})) were mathematically the identical within the RMSE precision considered, the same being true for Eq. (\ref{eq31}) and Eq. (\ref{eq36}), in both cases the curves for numerical and analytical solutions are superimposed in Fig. \ref{fig5}. 
 
\begin{figure}
\centering
 \includegraphics[width=0.75\textwidth]{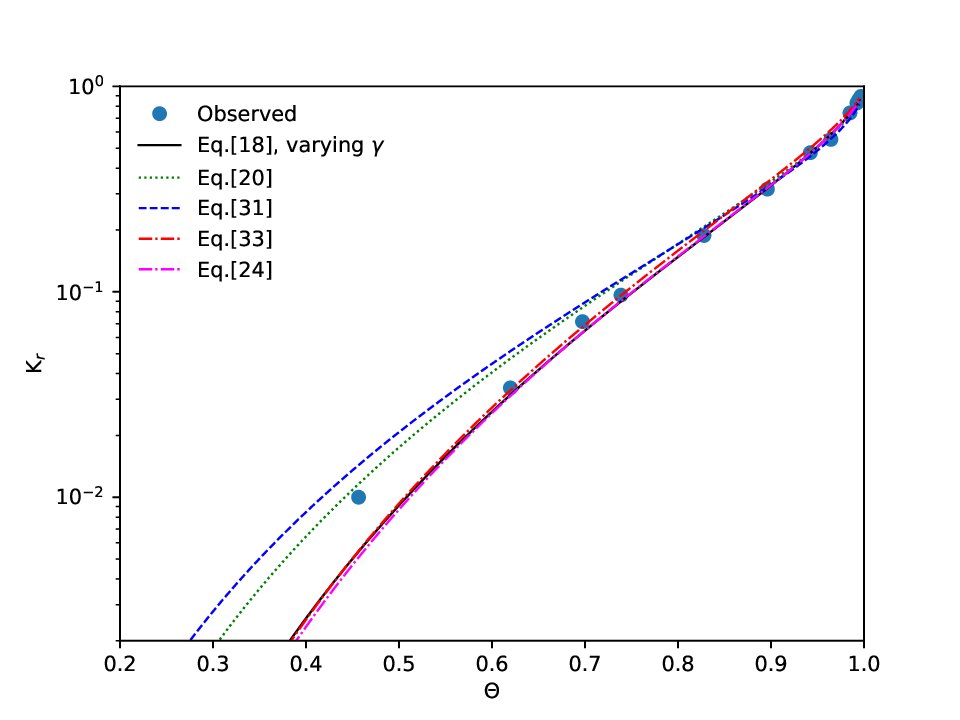}
\caption{Observed and estimated relative permeability as a function of relative saturation for selected models fit directly to permeability data.}
\label{fig6}
\end{figure}

 It is possible that the better accuracy for certain models was somewhat random as a variation of the combination of $m$ and $n$ will result on a best prediction of permeability. To test this, the permeability models were fit to observed permeability versus relative saturation data using nonlinear regression. In addition to the regular models, Eq. (\ref{eq18}) was also fit with varying $\gamma$. Better accuracy was obtained using Eq. (\ref{eq18}) with varying $\gamma$ $RMSE= 0.011813$, followed closely by Eq. (\ref{eq18}) $RMSE = 0.011967$ and Eq. (\ref{eq33}) $RMSE = 0.012132$, for Eq. (\ref{eq20}) and Eq. (\ref{eq31}) the $RMSE$ were $0.020459$ and $0.033136$, respectively (Figure \ref{fig6}). As before, the predictions from the analytical solutions exactly matched the special functions equations within the precision considered and were not included in this plot. By nonlinear regression theory it is expected the the general functions will outperform the restricted solutions by number of parameters alone, as the flexibility will be improved. However, the use of such functions could implicate in some gain in precision in hydrological models, which can have important consequences in large scale simulations. The performance of the models remain to be tested in independent $K_r(\Theta)$ datasets for a variety of soils and other porous media.

\section*{Declarations}

\begin{itemize}
\item Conflict of interests \\
The author could not uncover any conflict of interest.
\end{itemize}

\nocite{*}
\bibliography{sn-article-kbeta}

\end{document}